\documentclass[]{emulateapj}

\newcommand{\DDir}{\relax{D\kern-.7em{/}}}

\newcommand{\be}{\begin{equation}}
\newcommand{\ee}{\end{equation}}
\newcommand{\bea}{\begin{equation*}}
\newcommand{\eea}{\end{equation*}}

\newcommand{\nin}{\relax{\in\kern-.8em{/}}}

\usepackage{amsmath}
\begin{document}

\title{Closing the gap in the solutions of the strong explosion problem: An expansion of the family of second-type self-similar solutions}

\author{Doron Kushnir\altaffilmark{1}
and Eli Waxman\altaffilmark{1}} \altaffiltext{1}{Department of Particle Physics \& Astrophysics, Weizmann Institute of Science, Rehovot 76100, Israel}

\begin{abstract}

Shock waves driven by the release of energy at the center of a cold ideal gas sphere of initial density $\rho\propto r^{-\omega}$ approach a self-similar behavior, with velocity $\dot{R}\propto R^\delta$, as $R\rightarrow\infty$. For $\omega>3$ the solutions are of the second-type, i.e., $\delta$ is determined by the requirement that the flow should include a sonic point. No solution satisfying this requirement exists, however, in the $3\leq\omega\leq\omega_{g}(\gamma)$ ``gap'' ($\omega_{g}=3.26$ for adiabatic index $\gamma=5/3$). We argue that second-type solutions should not be required in general to include a sonic point. Rather, it is sufficient to require the existence of a characteristic line $r_c(t)$, such that the energy in the region $r_c(t)<r<R$ approaches a constant as $R\rightarrow\infty$, and an asymptotic solution given by the self-similar solution at $r_c(t)<r<R$ and deviating from it at $r<r_c$  may be constructed. The two requirements coincide for $\omega>\omega_g$ and the latter identifies $\delta=0$ solutions as the asymptotic solutions for $3\leq\omega\leq\omega_{g}$ \citep[as suggested by][]{Gruzinov03}. In these solutions, $r_c$ is a $C_0$ characteristic. It is difficult to check, using numerical solutions of the hydrodynamic equations, whether the flow indeed approaches a $\delta=0$ self-similar behavior as $R\rightarrow\infty$, due to the slow convergence to self-similarity for $\omega\sim3$. We show that in this case the flow may be described by a modified self-similar solution, $d\ln\dot{R}/d\ln R=\delta$ with slowly varying $\delta(R)$, $\eta\equiv d\delta/d\ln R\ll1$, and spatial profiles given by a sum of the self-similar solution corresponding to the instantaneous value of $\delta$ and a self-similar correction linear in $\eta$. The modified self-similar solutions provide an excellent approximation to numerical solutions obtained for $\omega\sim3$ at large $R$, with $\delta\rightarrow0$ (and $\eta\neq0$) for $3\leq\omega\leq\omega_{g}$.

\end{abstract}




\section{Introduction}
\label{sec:Introduction}

Self-similar solutions to the hydrodynamic equations describing adiabatic one-dimensional flows of an ideal gas are of interest for mainly two reasons. First, the nonlinear partial differential hydrodynamic equations are reduced for self-similar flows to ordinary differential equations, which greatly simplifies the mathematical problem of solving the equations and in certain cases allows one to find analytic solutions. Second, self-similar solutions often describe the limiting behavior approached asymptotically by flows which take place over a characteristic scale, $R$, which diverges or tends to zero \citep[see][for reviews]{SedovBook,ZeldovichRaizer,BarenblattBook}.

It is reasonable to assume that in the limit $R\rightarrow\infty(0)$ the flow becomes independent of any characteristic length scale. Using dimensional arguments, it is possible to show that in this case the flow fields must be of the self-similar form \citep{ZeldovichRaizer,WaxmanShvarts10}
\begin{equation}\label{eq:ss_scaling}
    u(r,t)=\dot{R}\xi U(\xi),\quad c(r,t)=\dot{R}\xi C(\xi),\quad \rho(r,t)=BR^\epsilon G(\xi),
\end{equation}
where $u$, $c$, and $\rho$ are the fluid velocity, sound speed, and density, respectively (the pressure is
given by $p=\rho c^{2}/\gamma$), and
\begin{equation}\label{eq:Rdot}
    \dot{R}=AR^\delta, \quad \xi(r,t)=r/R(t)
\end{equation}
\citep[for a somewhat different approach to self-similarity, based on Lie group methods, see][]{Coggeshall1986lgi,Coggeshall1991ash,Coggeshall1992gis}. For a self-similar solution of the form given in Equations~\eqref{eq:ss_scaling} and~\eqref{eq:Rdot}, the hydrodynamic equations, Equations~(\ref{eq:hydro_eq}), are replaced with a single ordinary differential equation, Equation~(\ref{eq:dUdC}),
\begin{equation}
\frac{dU}{dC}=\frac{\Delta_{1}(U,C)}{\Delta_{2}(U,C)},
\nonumber
\end{equation}
and one quadrature, Equation~(\ref{eq:quadrature}),
\begin{equation}
\frac{d\ln\xi}{dU}=\frac{\Delta(U,C)}{\Delta_{1}(U,C)}\qquad {\rm
or} \qquad \frac{d\ln\xi}{dC}=\frac{\Delta(U,C)}{\Delta_{2}(U,C)}.
\nonumber
\end{equation}
$\Delta$, $\Delta_1$, and $\Delta_2$ are given by Equations~(\ref{eq:deltas}).
As illustrated in Section~\ref{sec:gap} \citep[see also][]{Guderley42,Meyer-ter-Vehn82,WaxmanShvarts93}, many of the properties of self-similar flows may be inferred by analyzing the contours in the $(U,C)$-plane determined by Equation~(\ref{eq:dUdC}).

In this paper, we revisit the ``strong explosion problem'', which is one of the most familiar problems where asymptotic self-similarity is encountered. Consider the blast wave produced by the deposition of energy $E$ within a region of characteristic size $d$ at the center of an initially cold ($p=0$ at $r>d$) gas sphere with initial density $\rho_0=K r^{-\omega}$ (at $r>d$). As the shock radius $R$ diverges, we expect the flow to approach a self-similar solution of the form given by Equations~(\ref{eq:ss_scaling}) and~(\ref{eq:Rdot}). The asymptotic flow is described by the Sedov--von Neumann--Taylor (ST) solutions \citep{Sedov46,vonNeumann47,Taylor50} for $\omega<3$, and by the solutions derived by Waxman \& Shvrats \citep[WS;][]{WaxmanShvarts93,WaxmanShvarts10} for $\omega>3$. The ST solutions describe decelerating shocks ($\delta=(\omega-3)/2<0$) and are of the ``first-type'', where the similarity exponents, $\delta$ and $\epsilon$, are determined by dimensional considerations. The WS solutions describe accelerating shocks ($\delta>0$) and are of the ``second-type'', where the similarity exponents are determined by the condition that the solutions must pass through a singular point of Equation~(\ref{eq:dUdC}).

\begin{figure}
\epsscale{1.2} \plotone{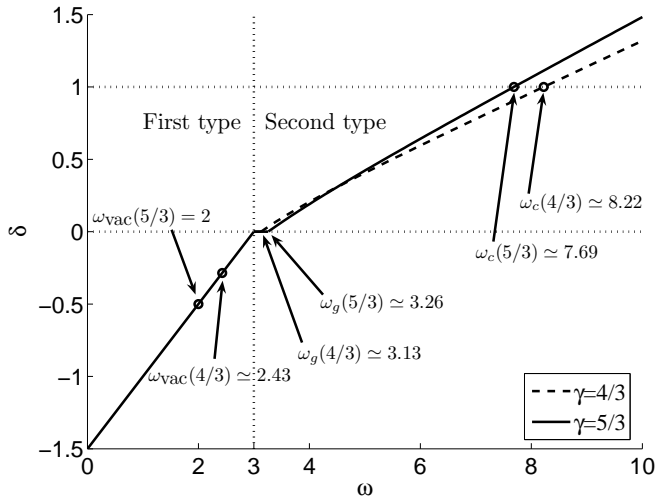} \caption{Self-similar exponent
$\delta$ as a function of $\omega$ for $\gamma=4/3$ (dashed line) and $\gamma=5/3$ (solid line).
\label{fig:delta-omega-full}}
\end{figure}

We revisit the strong explosion problem for several reasons. First, there exists a ``gap'' in the $(\gamma,\omega)$-plane, $3\leq\omega\leq\omega_{g}(\gamma)$, where neither the ST nor the WS solutions describe the asymptotic flow \citep[$\omega_{g}$ is increasing with $\gamma$, $\omega_{g}=3$ for $\gamma=1$ and $\omega_{g}\simeq3.26$ for adiabatic index $\gamma=5/3$;][]{WaxmanShvarts93}. Our first goal is to close this ``gap''. We argue in Section~\ref{sec:gap} that second-type solutions should not be required in general to include a sonic point, and that it is sufficient to require the existence of a characteristic line $r_c(t)$, such that the energy in the region $r_c(t)<r<R$ approaches a constant as $R\rightarrow\infty$. We show that the two requirements coincide for $\omega>\omega_g$ and that the latter requirement identifies $\delta=0$ solutions as the asymptotic solutions for $3\leq\omega\leq\omega_{g}$. This result is in agreement with that of \citet{Gruzinov03}, who suggested based on heuristic arguments that the $R\propto t$ solutions are the correct asymptotic solutions in the gap. As we explain in some detail at the end of Section~\ref{sec:gap_solutions}, the validity of the heuristic arguments is not obvious. We use a different reasoning, based on an extension of the analysis of \citet{WaxmanShvarts93}.

In Section~\ref{sec:numerical solution}, we compare the asymptotic, $R/d\gg1$, behavior of numerical solutions of the hydrodynamic equations, Equations~(\ref{eq:hydro_eq}), to that expected based on the $\delta=0$ self-similar solutions. We show that the convergence to self-similarity is very slow for $\omega\sim3$. Hence, it is difficult to check using numerical solutions whether the flow indeed approaches a $\delta=0$ self-similar behavior as $R\rightarrow\infty$. We show in Section~\ref{sec:mod_slfsim} that in this case the flow may be described by a modified self-similar solution, $d\ln\dot{R}/d\ln R=\delta$ with slowly varying $\delta(R)$, $\eta\equiv d\delta/d\ln R\ll1$, and spatial profiles given by a sum of the self-similar solution corresponding to the instantaneous value of $\delta$ and a self-similar correction linear in $\eta$. The modified self-similar solutions provide an excellent approximation to numerical solutions obtained for $\omega\sim3$ at large $R$, with $\delta\rightarrow0$ (and $\eta\neq0$) for $3\leq\omega\leq\omega_{g}$.

The second reason for revisiting the strong explosion problem is that it is of general methodological interest. It demonstrates transitions between (mathematically and physically) different types of solutions as the value of $\omega$ changes, as illustrated in Figure~\ref{fig:delta-omega-full}: as the value of $\omega$ increases above $\omega=3$, a transition occurs between first-type solutions (at $\omega<3$) and second-type solutions \citep[at $\omega>3$;][]{WaxmanShvarts93}; as the value of $\omega$ increases above $\omega=\omega_c(\gamma)$ ($\omega_c\sim8$ for $4/3<\gamma<5/3$), the $\omega<\omega_c$ power-law solutions ($R\propto t^{1/(1-\delta)}$, $\delta<1$) are replaced with exponential solutions ($R\propto e^{t/\tau}$, $\delta=1$) at $\omega=\omega_c$ and with solutions diverging in finite time ($R\propto (-t)^{1/(\delta-1)}$, $\delta>1$) at $\omega>\omega_c$ \citep{WaxmanShvarts10}. We show here that the strong explosion problem also exhibits a transition between two sub-types of second-type solutions at $\omega=\omega_{g}(\gamma)$. We argue in Section~\ref{sec:Discussion} that, based on the results presented in this paper, the definition of the two types of self-similar solutions should be somewhat modified and that the family of asymptotic second-type solutions should be expanded.

Finally, we note that the propagation of shock waves in steep density gradients is of interest in a wide variety of astrophysical contexts \citep[e.g.][and references therein]{OstrikerMcKee88,KooMcKee90}, such as supernova explosions \citep[e.g.][and references therein]{MM99}. It is worth noting that self-similar solutions for shock propagation in power-law density profiles, $\rho\propto r^{-\omega}$, are useful for describing shock propagation in more general density profiles \citep[e.g.][]{MM99,OrenSari09}, as well as for the general study of shock wave stability \citep[e.g.][]{Goodman90,Chevalier90,Kushnir2005tsd,SWS00}.


\section{Self-similar solutions in the ``gap'' region}
\label{sec:gap}

As explained in Section~\ref{sec:Introduction}, we expect the strong explosion flow to approach a self-similar behavior, of the form given by Equations~(\ref{eq:ss_scaling}) and~(\ref{eq:Rdot}), as $R$ diverges. Since for strong shocks the density just behind the shock wave is a constant factor, $(\gamma+1)/(\gamma-1)$, times the density just ahead of the shock, we must have $\epsilon=-\omega$, and we may choose $B=K$. With this normalization, the Rankine--Hugoniot relations at the shock front determine the boundary conditions for the self-similar solutions to be \citep[e.g.][]{ZeldovichRaizer}
\begin{equation}\label{eq:shock_boundary}
    U(1)=\frac{2}{\gamma+1},\quad C(1)=\frac{\sqrt{2\gamma(\gamma-1)}}{\gamma+1}, \quad G(1)=\frac{\gamma+1}{\gamma-1}.
\end{equation}
The only parameter of the self-similar solution that remains to be determined is $\delta$.

\subsection{ST and WS solutions}
\label{sec:ST_WS}

\begin{figure}
\epsscale{1} \plotone{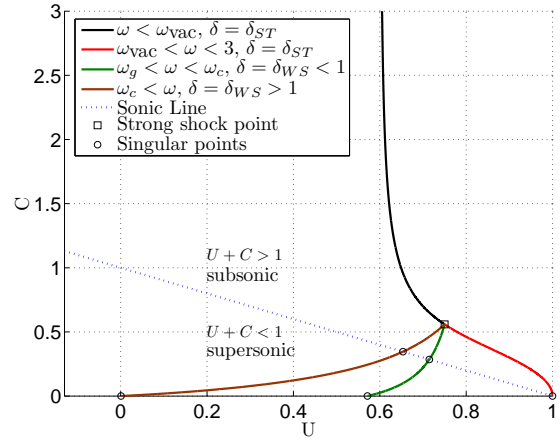} \caption{Different types of the $C(U)$ curves of the solutions of the strong explosion problem, obtained for $\omega$ values in the regimes $\omega<\omega_{\rm vac}=(7-\gamma)/(\gamma+1)$, $\omega_{\rm vac}<\omega<3$, $\omega_g<\omega<\omega_c$, and $\omega_c<\omega$. Also shown is the sonic line, $U+C=1$. The square denotes the strong shock point, Equations~(\ref{eq:shock_boundary}), and the circles denote the singular points $(U,C)=(0,0)$, $(U,C)=(1-\delta,0)$, and $(U,C)=(1,0)$.
\label{fig:UC_curves}}
\end{figure}

The self-similar solution, given by Equations~(\ref{eq:ss_scaling}) and~(\ref{eq:Rdot}), depends on two independent dimensional constants, $A$ and $B=K$. In the ST analysis, it is assumed that the second dimensional constant, in addition to $K$, that determines the self-similar solution is $E$. In this case, dimensional considerations imply $R\propto (Et^2/K)^{1/(5-\omega)}$, i.e.,
\begin{equation}\label{eq:delta_ST}
    \delta=\delta_{\textrm{ST}}\equiv\frac{\omega-3}{2}.
\end{equation}
For $\omega<3$ we have $\delta<0$, i.e., decelerating blast waves. The flow properties are qualitatively different in the regimes $\omega<\omega_{\textrm{vac}}\equiv(7-\gamma)/(\gamma+1)$ and $\omega_{\textrm{vac}}<\omega<3$ (see Figure~\ref{fig:UC_curves}). For $\omega<\omega_{\textrm{vac}}$, $U$ tends to $1/\gamma$ and $C$ tends to infinity as $\xi$ tends to zero. For
$\omega_{\textrm{vac}}<\omega<3$, the self-similar solution contains an ``evacuated'' region: there exists some finite $\xi_{\rm in}>0$, such that the spatial region $0<\xi<\xi_{\rm in}$ is evacuated ($\rho=0$). The self-similar solution describes the flow for
$\xi_{\rm in}\le \xi\le 1$ and is matched to the evacuated region, $\xi<\xi_{\rm in}$, by a weak discontinuity, which lies at $\xi=\xi_{\rm in}$. In this case, $U$ tends to 1 and $C$ tends to 0 as $\xi$ tends to $\xi_{\rm in}$. A detailed discussion of the ST solutions is given by \citet{Korobeinikov1991ppb} and \citet{Book1994}.

As explained in detail in \citep{WaxmanShvarts93}, the ST solutions are the correct asymptotic solutions only for $\omega<3$, for which $\delta<0$. For larger values of $\omega$ the mass and energy contained in the self-similar solution are infinite, reflecting the fact that the initial gas mass at $r>d$ diverges for $d\rightarrow0$. It was therefore suggested by \citet{WaxmanShvarts93} that for $\omega>3$ the asymptotic solution is given by a self-similar solution only over part of the $(\xi,R)$-plane, bounded by $\xi=1$ and $\xi_c(R)<1$, and by a different solution at $0<\xi<\xi_c(R)$.

\begin{figure}
\epsscale{1} \plotone{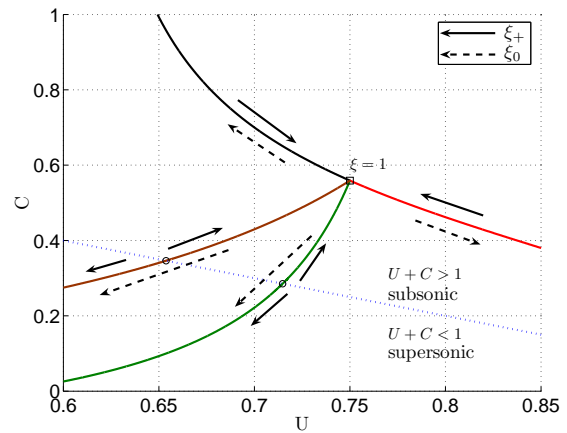} \caption{Zoomed version of Figure~\ref{fig:UC_curves} (using the same line types). Solid arrows indicate the propagation direction of $C_{+}$ characteristics and dashed arrows indicate the propagation direction of $C_{0}$ characteristics.
\label{fig:UC_curves_zoom}}
\end{figure}

Since such a solution includes a contact or a weak discontinuity at $\xi_c(R)$, $\xi_c(R)$ must coincide with a characteristic of the self-similar solution. For the self-similar flow, the characteristic lines
\begin{equation}\label{eq:characteristics}
    C_0:\frac{dr_0}{dt}=u,\quad C_\pm: \frac{dr_\pm}{dt}=u\pm c
\end{equation}
are given by
\begin{eqnarray}\label{eq:slfsim_char}
C_{0}&:&\frac{d \ln \xi_{0}}{d \ln R}=U(\xi_{0})-1,\nonumber \\
C_{\pm}&:&\frac{d \ln \xi_{\pm}}{d \ln R}=U(\xi_{\pm})\pm
C(\xi_{\pm})-1.
\end{eqnarray}
The directions in which the different characteristics propagate are illustrated in Figure~\ref{fig:UC_curves_zoom}. The physical interpretation of the behavior illustrated in this figure is as follows.
The flow just behind the shock is always subsonic: the shock-front point $(U,C)=(U(1),C(1))$ lies above the ``sonic line'' $U+C=1$, i.e., $U(1)+C(1)>1$, which implies that $C_+$ characteristics emerging from points just behind the shock always overtake it. $C_+$ characteristics that do not overtake the shock exist only if the self-similar solution crosses the $U+C=1$ line in the $(U,C)$-plane into the region where $U+C<1$. $C_{0}$ characteristics, however, never overtake the shock and propagate away from it (in $\xi$ space).

Requiring $\xi_c$ to coincide with a $C_+$ characteristic, that does not overtake the shock, implies therefore that the solution must cross the $U+C=1$ line. Since $\Delta=0$ along the sonic line, Equation~(\ref{eq:quadrature}) imply that a physical solution must cross the sonic line at a singular point $\Delta_1=\Delta_2=0$, as otherwise $U(\xi)$ and/or $C(\xi)$ are not single valued. This requirement determines the correct value of $\delta$ for the $\omega>3$ asymptotic solutions.

The self-similar solutions obtained in this way for $\omega>3$ were analyzed in detail in \citep{WaxmanShvarts93,WaxmanShvarts10}. They describe accelerating blast waves, with $\delta>0$. As $\xi\rightarrow0$ the $C(U)$ curves of these solutions approach the singular point $(U,C)=(1-\delta,0)$ for $\delta<1$, and the singular point $(U,C)=(0,0)$ for $\delta\ge1$ (see Figure~\ref{fig:UC_curves}). Analyzing the behavior of the solutions near these singular points, it was shown that although the mass and energy contained in the self-similar solution are infinite, the mass and energy contained within the region $\xi_c(R)<\xi<1$ approach finite values as $R\rightarrow\infty$, for any $C_+$ characteristic which satisfies $\xi_c(R)\rightarrow0$ as $R\rightarrow\infty$. Moreover, it was shown that $\xi_c(R)R\propto t$ as $R\rightarrow\infty$, implying that the asymptotic flow within the region $0<r<\xi_c(R)R$ is described by the self-similar solution of expansion into vacuum. Finally, it was demonstrated by numerical calculations that the asymptotic behavior described above is indeed approached for $R/d\gg1$ \citep{WaxmanShvarts93,WaxmanShvarts10}.

The self-similar solutions derived by \citet{WaxmanShvarts93} exist only for $\omega>\omega_g(\gamma)>3$. Within the range $3<\omega<\omega_g(\gamma)$, there is no value of $\delta$ for which the $C(U)$ curve crosses the sonic line at a singular point ($\omega_g(\gamma)>3$ is increasing with $\gamma$, with $\omega_g=3.26$ for $\gamma=5/3$ and  $\omega_g\rightarrow3$ for $\gamma\rightarrow1$, see Figure~\ref{fig:delta-omega-full}). Since the ST solutions provide the correct asymptotic solutions only for $\omega<3$, the asymptotic behavior within the (narrow) range of $3<\omega<\omega_g(\gamma)$ is not described by either of the two types of solutions.

\subsection{The asymptotic solutions for $\omega$ values within the ``gap''}
\label{sec:gap_solutions}

As explained in the previous section, the divergence of the energy enclosed in the self-similar solutions for $\omega>3$ suggests that the asymptotic solution should be composed of a self-similar solution describing the flow at $\xi_c(R)<\xi<1$, matched to a different solution at $0<\xi<\xi_c(R)$. In general, $\xi_c$ may be a characteristic line of any type. In the analysis of \citep{WaxmanShvarts93}, it was assumed that $\xi_c$ is a $C_+$ characteristic. This was mainly motivated by the fact that requiring the existence of a $C_+$ characteristic that does not overtake the shock front as $R\rightarrow\infty$ is equivalent to requiring that the solution passes through a sonic point, and it is commonly accepted that the similarity exponents of a second-type solution are determined by the requirement that the solution passes through such a singular point. We argue here that in order to determine the similarity exponents it is sufficient to require the existence of any characteristic that does not overtake the shock and for which the energy contained in the self-similar part of the flow,  $\xi_c(R)<\xi<1$, does not diverge as $R\rightarrow\infty$. Examining $\omega$ values below, within, and above the ``gap'' we show that for each value of $\omega$ there is only one value of $\delta$ which yields a valid solution: a solution that does not cross the sonic line at a non-singular point and that either contains a finite energy or a characteristic line satisfying the conditions described above.

The energy contained in the  $\xi_c(R)<\xi<1$ region of a self-similar solution is
\begin{eqnarray}\label{eq:E_ss}
E_{s}(R)&\equiv&\int_{\xi_c(R)R}^Rdr 4\pi r^2\left(\frac{1}{2}\rho u^2+\frac{1}{\gamma-1}p\right) \nonumber\\
&=&4\pi R^{3-\omega+2\delta}A^2 K
\left\{I_{k}[\xi_{c}(R)]+I_{i}[\xi_{c}(R)]\right\},
\end{eqnarray}
with
\begin{equation}\label{eq:Ik def}
I_{k}(\xi)=\int\limits_{\xi}^{1}d\xi'\xi'^{4}G\frac{1}{2}U^{2}, \quad
I_{i}(\xi)=\int\limits_{\xi}^{1}d\xi'\xi'^{4}G\frac{1}{\gamma(\gamma-1)}C^{2}.
\end{equation}
The  $I_{k}$ and $I_{i}$ terms give the kinetic and internal energy of the gas, respectively. In order for $E_s(R)$ not to diverge as $R\rightarrow\infty$ we must have
\begin{equation}\label{eq:delta_lim}
    \delta\le\delta_{\textrm{ST}}=\frac{\omega-3}{2}.
\end{equation}
For $\delta>\delta_{\textrm{ST}}$ the energy contained in any $\xi_1<\xi<\xi_2$ region of the self-similar solution diverges since $R^{3-\omega+2\delta}$ diverges as $R\rightarrow\infty$. For the WS solutions, $\delta<\delta_{\textrm{ST}}$ and $I_{k}(\xi)$ diverges as $\xi\rightarrow0$ in such a manner that the product $R^{3-\omega+2\delta}I_{k}[\xi_{c}(R)]$ tends to a constant as $R$ diverges.

\begin{deluxetable*}{cccc}
\tablecaption{Determining $\delta$ Based on the Properties of the Self-similar Solutions \label{tbl:tbl1}}
\tablewidth{0pt} \tablehead{ \colhead{Properties of the self-similar solutions} & \colhead{$\omega_{\textrm{vac}}<\omega<3$} & \colhead{$3<\omega<\omega_{g}(\gamma)$} & \colhead{$\omega_{g}(\gamma)<\omega$}} \startdata $C(U)$ crosses the sonic line at a non-singular point & $\delta<\delta_{\textrm{ST}}$ & $\delta<0$ & $\delta<\delta_{\textrm{WS}}$ \\   \hline
$C(U)$ crosses the sonic line at a singular point and & & & \\
$E_s(R)$ does not diverge as $R\rightarrow\infty$ & ... & ... & $\delta=\delta_{\textrm{WS}}$ \\   \hline
$C(U)$ terminates at $(U,C)=(1,0)$ and & & & \\ $E_s(R)$ does not diverge as $R\rightarrow\infty$ & $\delta=\delta_{\textrm{ST}}$ & $\delta=0$ & ... \\  \hline
$C(U)$ terminates at $(U,C)=(1,0)$ and & & &  \\ $E_s(R)$ diverges as $R\rightarrow\infty$ & $\delta>\delta_{\textrm{ST}}$ & $\delta>0$ &  $\delta>\delta_{\textrm{WS}}$ \\
\enddata
\label{table}
\end{deluxetable*}

Let us consider the $C(U)$ curves, the behavior of characteristic lines, and the behavior of $E_s(R)$ for self-similar solutions with $\epsilon=-\omega$, that satisfy the strong shock boundary conditions, Equations~(\ref{eq:shock_boundary}). We examine $\omega$ values below the gap, $\omega_{\textrm{vac}}<\omega<3$, within the gap, and above the gap, $\omega_g<\omega<\omega_c$. Table~\ref{table} summarizes the relevant properties of the solutions obtained at the different $\omega$ regions for different values of $\delta$. The derivation of these properties is described below.

First, we consider the properties of the $C(U)$ curves by numerically investigating the solutions of Equation~(\ref{eq:dUdC}). We find that for $\delta$ values smaller than a critical value, $\delta_*(\omega)$, the $C(U)$ curves starting at the strong shock point cross the sonic line at a non-singular point. We find that $\delta_*=\delta_{\textrm{ST}}$ for $\omega_{\textrm{vac}}<\omega<3$, $\delta_*=0$ for $3<\omega<\omega_g$, and $\delta_*=\delta_{\textrm{WS}}$ for $\omega_g<\omega<\omega_c$. As explained above, the self-similar solutions obtained for $\delta<\delta_*(\omega)$ are not physical. For $\omega_g<\omega$ and $\delta=\delta_*(\omega)=\delta_{\textrm{WS}}$, the $C(U)$ curves cross the sonic line at a singular point. Finally, for other values of $\delta$, $\delta\ge\delta_*(\omega)$ for $3<\omega<\omega_g$ and $\delta>\delta_*(\omega)$ for $\omega_g<\omega$, the $C(U)$ curves do not cross the sonic line and terminate at the singular point $(U,C)=(1,0)$.

Next, we examine the energy contained in the self-similar solution. For $\omega<3$, the only physical solution is that obtained for $\delta=\delta_{\textrm{ST}}$: $\delta$ must satisfy Equation~(\ref{eq:delta_lim}), $\delta\le\delta_{\textrm{ST}}$, in order for the energy in any $\xi_1<\xi<\xi_2$ part of the solution not to diverge, and the $C(U)$ curve crosses the sonic line at a non-singular point for $\delta<\delta_{\textrm{ST}}$. In order to examine the energy content of solutions in the $(3<\omega<\omega_g,\delta\ge0)$ and $(\omega_g<\omega,\delta>\delta_{\textrm{WS}})$ regions we need to analyze the behavior of the solutions near the singular point $(U,C)=(1,0)$. The $C(U)$ curves of the solutions in these regions of $(\omega,\delta)$ lie above the sonic line, which implies that all $C_+$ characteristics emerging behind the shock overtake the shock. $C_0$ and $C_-$ characteristics, on the other hand, move along the $C(U)$ curve toward the $(U,C)=(1,0)$ singular point (see Figures~\ref{fig:UC_curves} and~\ref{fig:UC_curves_zoom}).

Defining $f=1-U$, Equation~(\ref{eq:dUdC}) is given, to leading orders in $f$ and $C$, by
\begin{equation}\label{eq:fC near edge}
\frac{d\ln f}{d\ln C}=
\begin{cases}
    {\frac{\delta f+[3-(\omega-2\delta)/\gamma]C^2}
{-\delta f(\gamma-1)/2+\{[(\gamma-1)\omega+2\delta]/(2\gamma)\}C^2}} & \text{for $\delta\neq0$,} \\
    {\frac{f^2-(3-\omega/\gamma)C^2}
{(\gamma-1)f^2-[(\gamma-1)\omega/(2\gamma)]C^2}} & \text{for $\delta=0$.} \
\end{cases}
\end{equation}

Let us consider first the $\delta>0$ case. Regardless of whether $f$ approaches 0 faster or slower than $C^2$, Equation~(\ref{eq:fC near edge}) implies that in this limit
\begin{equation}\label{eq:f on C near edge}
\lim_{f\to 0}\frac{d\ln f}{d\ln C}=\nu,
\end{equation}
where $\nu$ is some constant. Assuming that $f$ approaches zero slower than $C^2$, i.e., that $\nu<2$, leads to contradictions since Equation~(\ref{eq:fC near edge}) gives $\nu=-2/(\gamma-1)<0$. Assuming that  $f$ approaches zero faster than $C^2$, i.e., that $\nu>2$, Equation~(\ref{eq:fC near edge}) gives $\nu=(6\gamma-2\omega+4\delta)/[(\gamma-1)\omega+2\delta]$, for which $\nu>2$ implies $\omega<3$. Thus, for $\omega>3$ and $\delta>0$ we must have $\nu=2$, i.e., $f\propto C^2$, and Equation~(\ref{eq:fC near edge}) gives
\begin{equation}\label{eq:f_C1}
    f=\frac{\omega-3}{\gamma\delta} C^2.
\end{equation}
Using this result, the quadrature, Equation~(\ref{eq:quadrature}), gives
\begin{equation}\label{eq:f_xi_1}
    f=\frac{3(\gamma-1)+2\delta}{\gamma}\ln\left(\frac{\xi}{\xi_{\rm in}}\right),
\end{equation}
i.e., the singular point $(U,C)=(1,0)$ is approached for finite $\xi_{\rm in}>0$. Using these results and Equation~(\ref{eq:G}) we find
\begin{equation}\label{eq:G1}
    G\propto f^{-(\gamma\omega+2\delta-3)/[3(\gamma-1)+2\delta]}.
\end{equation}

We may now determine the $R$ dependence of the energy $E_s(R)$, contained within the $\xi_0(R)<\xi<1$ region of the self-similar solution, where $\xi_0$ is a $C_0$ characteristic. Equation~(\ref{eq:slfsim_char}) may be solved, using Equation~(\ref{eq:f_xi_1}), to give
\begin{equation}\label{eq:C0_1}
    \ln\left(\frac{\xi_0}{\xi_{\rm in}}\right)\propto R^{-[3(\gamma-1)+2\delta]/\gamma}
\end{equation}
for the evolution of $C_0$ characteristics. Using Equations~(\ref{eq:f_xi_1}) and~(\ref{eq:G1}), we find that the kinetic energy integral, $I_k(\xi)$ given in Equation~(\ref{eq:Ik def}), diverges in the limit $\xi\rightarrow\xi_{\rm in}$ as
\begin{eqnarray}\label{eq:Ik_eta1}
    I_k(\xi)&\propto& f(\xi)^{1-(\gamma\omega+2\delta-3)/[3(\gamma-1)+2\delta]}\nonumber\\
    &\propto& \left[\ln\left(\frac{\xi}{\xi_{\rm in}}\right)\right]^{-\gamma(\omega-3)/[3(\gamma-1)+2\delta]}.
\end{eqnarray}
Using Equation~(\ref{eq:C0_1}) we then find
\begin{eqnarray}\label{eq:Ik_C01}
    I_k[\xi_0(R)]&\propto& R^{(\omega-3)},
\end{eqnarray}
which, using Equation~(\ref{eq:E_ss}), implies that the energy diverges (for $\delta>0$) as $E_s\propto R^{2\delta}$.

This divergence of energy implies that the $\delta=\delta_{\textrm{WS}}>0$ solutions are the only physical solutions for $\omega>\omega_g$. These solutions cross the sonic line at a singular point and terminate at $(U,C)=(1-\delta,C=0)$ with finite energy $E_s$, while the $\delta>\delta_{\textrm{WS}}>0$ solutions terminate at $(U,C)=(1,0)$ with diverging energy $E_s$. The divergence also implies that the $\delta=0$ solutions are the only solutions that may be physical solutions within the gap. Let us consider therefore the behavior of the $\delta=0$ solutions next.

The solution of Equation~(\ref{eq:fC near edge}) for $\delta=0$ must also be of the form given by Equation~(\ref{eq:f on C near edge}). Assuming $f$ tends to 0 slower than $C$, i.e., $\nu<1$, leads to a contradiction since Equation~(\ref{eq:fC near edge}) gives $\nu=1/(\gamma-1)>1$. Therefore, $\nu$ must satisfy $\nu\ge1$. For $\nu>1$, Equation~(\ref{eq:fC near edge}) gives
\begin{equation}\label{eq:f_C}
    \nu=\frac{6\gamma-2\omega}{(\gamma-1)\omega},
\end{equation}
which satisfies $\nu>1$ for $\omega<6\gamma/(\gamma+1)$. For $\nu=1$, Equation~(\ref{eq:fC near edge}) gives
\begin{equation}\label{eq:f_C_nu1}
    f^2=\frac{6\gamma-(\gamma+1)\omega}{2\gamma(2-\gamma)} C^2.
\end{equation}
The solution of the quadrature, Equation~(\ref{eq:quadrature}), gives
\begin{equation}\label{eq:f_xi}
    f=\theta\ln\left(\frac{\xi}{\xi_{\rm in}}\right),\quad\theta=
    \left\{
      \begin{array}{ll}
        \left(3-\frac{\omega}{\gamma}\right), & \hbox{$\nu>1$ ;} \\
        \frac{3(\gamma-1)}{\gamma+1}, & \hbox{$\nu=1$,}
      \end{array}
    \right.
\end{equation}
which implies
\begin{equation}\label{eq:C0}
    \ln\left(\frac{\xi_0}{\xi_{\rm in}}\right)\propto R^{-\theta}.
\end{equation}
Using these results and Equation~(\ref{eq:G}) we find
\begin{equation}\label{eq:Geta}
    G\propto f^{-\mu}, \quad \mu=
    \left\{
      \begin{array}{ll}
        (\gamma-1)\omega/(3\gamma-\omega), & \hbox{$\nu>1$ ;} \\
        ((\gamma+1)\omega-6)/3(\gamma-1), & \hbox{$\nu=1$,}
      \end{array}
    \right.
\end{equation}
which implies that $I_k(\xi)$ diverges in the limit $\xi\rightarrow\xi_{\rm in}$ as $I_k(\xi)\propto f(\xi)^{1-\mu}$, which gives
\begin{eqnarray}\label{eq:Ik_C0}
    I_k(\xi_0)&\propto& R^{\theta(\mu-1)}=R^{\omega-3}.
\end{eqnarray}
Numerical integration of Equations~(\ref{eq:dUdC}) and~(\ref{eq:quadrature}) shows that solutions starting at the strong shock point, Equations~(\ref{eq:shock_boundary}), approach the $(U,C)=(1,0)$ singular point along a $\nu>1$ curve. The asymptotic behavior of $I_k$ is the same, as we find here, for both $\nu=1$ and $\nu>1$.

Using Equation~(\ref{eq:E_ss}), we find that the kinetic energy part of $E_s$ approaches a finite, non-zero, constant as $R$ diverges (it is straightforward to verify that the internal energy part of $E_s$ vanishes in the limit $R\rightarrow\infty$). The fact that the kinetic energy approaches a constant also implies that the mass contained within $\xi_0(R)<\xi<1$ approaches a finite non-zero constant, since the velocity of each fluid element, $\xi_0\dot{R}$, approaches a constant, $\xi_{\rm in}\dot{R}$. Thus, the $\delta=0$ solutions satisfy the requirement for the existence of a characteristic $\xi_c(R)$ that does not overtake the shock, and for which the energy contained in the self-similar part of the flow,  $\xi_c(R)<\xi<1$, does not diverge as $R\rightarrow\infty$.

As pointed out in Section~\ref{sec:Introduction}, it was suggested by \citet{Gruzinov03} that the asymptotic solutions in the gap are the $\delta=0$ solutions. The justification given there is based on the argument that the non-self-similar part of the asymptotic flow, which must exist since the self-similar solution contains infinite energy, acts as an infinite mass piston, which must move at a constant speed and may support the flow ahead of it. As we show here, the mass and energy contained in the self-similar and non-self-similar parts of the flow are both finite and may be comparable (this is also the case for the WS solutions obtained for $\omega>\omega_g$). Thus, the validity of the heuristic argument given in \citep{Gruzinov03} is not obvious.



\section{Slow convergence to self-similarity: Numerical results and modified self-similar solutions}
\label{sec:slowly converging}

\subsection{Comparison to numerical solutions}
\label{sec:numerical solution}

We present in this section a comparison between the $\delta=0$ self-similar solutions and numerical solutions of the flow
equations obtained for $\omega$ values within the gap. We have numerically calculated the propagation of a strong spherical shock wave through an ideal gas using a Lagrangian scheme with total energy conservation \citep[e.g.][]{Caramana98}. Shock waves are described in our calculations using a von Neumann artificial viscosity, implemented as a pressure term in all cells with negative difference between the nodes' velocities, $\Delta u<0$,
\begin{equation}\label{eq:art visc}
q=\Delta u\left(x_{q}\Delta u-x_{l}c\right),
\end{equation}
with $x_{q}=4$ and $x_{l}=0.1$. The initial conditions used are zero velocity everywhere, constant density and pressure at $r<d$, relatively small pressure at $r>d$ (see below), and a density profile proportional to $r^{-\omega}$ at $r>d$. The initial mesh spacing was uniform, $\Delta r=d$, i.e., the pressure was high only in the innermost cell. We chose the density of the innermost cell such that its mass is $10$ times higher than its neighbor's, and we chose its pressure to be $10^{6}$ times higher than its neighbor's. The pressure at the rest of the cells was chosen such that the outgoing shock wave is always strong. The mesh included $10^{5}$ cells, i.e., we were able to calculate explosions up to a radius of $R/d\simeq10^{5}$.

\begin{figure}
\epsscale{1} \plotone{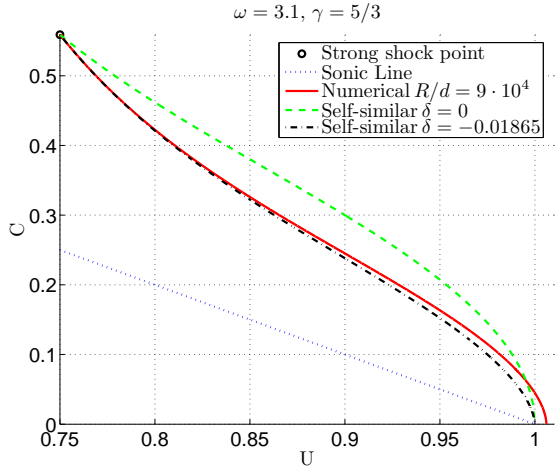} \caption{$C(U)$ curve for $\omega=3.1,\, \gamma=5/3$. Shown are the numerical solution at $R/d=9\times10^{4}$ and two self-similar solutions, corresponding to $\delta=0$ and $\delta=-0.01865$. Note that the $C(U)$ curve of the $\delta=-0.01865$ self-similar solution crosses the sonic line (at a non-singular point). This is not visible in the figure since the crossing takes place very close to $(U,C)=(1,0)$.
\label{CU}}
\end{figure}

\begin{figure}
\epsscale{1} \plotone{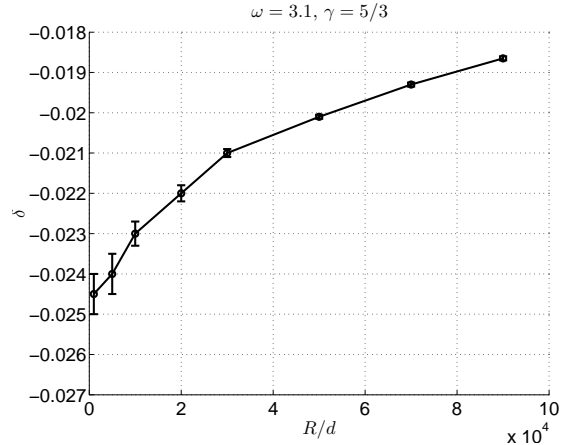} \caption{$\delta(R)$ determined from a numerical simulation, using the zero slope method (see the text), for an explosion with $\omega=3.1,\, \gamma=5/3$. The error bars are an estimate of the accuracy of the determination of $\delta$ using this method. The accuracy is better for larger values of $R/d$, where the flow behind the shock is better resolved.
\label{deltaofR}}
\end{figure}

We would like to examine the behavior of $\delta(R)\equiv d\ln \dot{R}/d\ln R=R\ddot{R}/\dot{R}^2$. Since $\delta$
depends on $\ddot{R}$ and the numerically determined value of $R$ is noisy due to the finite resolution, a derivation of $\delta$ by a direct differentiation of $R$ is not accurate enough for our study. In order to overcome this problem, we derive $\delta$ from the numerical spatial profiles: we choose $\delta$ as the value for which the difference between the numerical profiles and the self-similar profiles, determined by Equations~(\ref{eq:dUdC})--(\ref{eq:G}), has a zero slope at the shock front (this ensures that the acceleration of the shock wave, which is determined by the spatial profiles in the vicinity of the shock front, is the same for the self-similar and numerical solutions).
In what follows, we give some details regarding this method for the determination of $\delta$, which we term the ``zero slope method''.

We examine the difference $f(\xi)$ between the self-similar profiles ($U$, $C$, and $P$), obtained for a chosen value of $\delta$, and the profiles obtained in the numerical simulations at some $R/d$. For the comparison of the self-similar and numerical profiles we consider the radial range $0.999>\xi=r/R>0.9995$. The upper limit of this range is set by requiring that the oscillations behind the shock, caused by the artificial viscosity, are damped considerably, while the lower limit is chosen to ensure that deviations from the self-similar solutions are small (see Section~\ref{sec:mod_slfsim}). In order to determine whether or not the difference $f(\xi)$ is consistent with a zero slope, i.e., with $f'=0$, it is insufficient to use a simple linear fit for $f(\xi)$, since oscillations in the numerical profile caused by the artificial viscosity are not random and cannot be neglected. We define therefore $\bar{f}=(f-\mu(f))/\sigma(f)$ and $\bar{\xi}=(\xi-\mu(\xi))/\sigma(\xi)$, where $\mu$ and $\sigma$ stand for mean and standard deviation over all numerical cells in the range, and examine the number of points for which $\bar{\xi}>1$ and $\bar{f}>0(<0)$, denoted $N_{++(-)}$, and the number of points for which $\bar{\xi}<1$ and $\bar{f}>0(<0)$, denoted by $N_{-+(-)}$. In the absence of numerical inaccuracies, the ratios $r_{+(-)}=N_{+(-)+}/N_{+(-)-}$ should all equal unity for $f'=0$. In order to allow for numerical inaccuracies, we consider $f'$ to be consistent with 0 for $0.1<r_{+(-)}<10$, and determine the range of allowed values of $\delta$ as the range for which $0.1<r_{+(-)}<10$. Since $r_{+(-)}$ (or $1/r_{+(-)}$) grow rapidly as $\delta$ is modified, the range of allowed values of $\delta$ is not sensitive to the exact choice of the allowed range of $r_{+(-)}$.

In Figure~\ref{CU}, we compare the numerical $C(U)$ curve obtained at $R/d=9\times10^{4}$, for an explosion with $\omega=3.1,\, \gamma=5/3$, with the self-similar curve obtained for the value of $\delta$ determined by the zero slope method, $\delta=-0.01865$. The numerical curve and the self-similar one are very close near the shock front and show small discrepancy far from the shock. Similar results are obtained for other values of $R/d$.

$\delta(R)$ determined by the method describe above is shown in Figure~\ref{deltaofR} for an explosion with $\gamma=5/3,\,\omega=3.1$. The error bars are an estimate of the accuracy of the determination of $\delta$ using this method. It is apparent that the convergence of the inferred value of $\delta$ is very slow, and that it has not converged for a very large value of $R/d$, $R/d\approx10^{5}$. It is difficult to determine, based on the simulation, whether or not $\delta$ approaches 0 for $R/d\rightarrow\infty$: the rate of change of $\delta$ obtained at $R/d\approx10^{5}$, $d\delta/d\ln(R)\approx3\times10^{-3}$ implies that an increase in $R/d$ by a factor of $10$ will modify the inferred value of $\delta$ from $\delta\approx-0.019$ to only give $\delta\approx-0.014$.

A comment is in order regarding the convergence of our numerical solutions. We have checked convergence by examining the modification of the inferred value of $\delta$ at fixed $R/d$, obtained when the number of cells within the region of high initial pressure is increased, i.e., choosing $\Delta r/d<1$ (and keeping a uniform initial grid spacing). The results presented in this section for $\delta$ at large $R/d$ using $\Delta r/d=1$ are converged to a few percent. For example, for the $\gamma=5/3,\,\omega=3.1$ case, increasing the number of cells by a factor of $2$, i.e., using $\Delta r/d=1/2$, changes the inferred values of $\delta$ by less than a $5\%$.

Slow convergence of the numerical solutions to an asymptotic self-similar behavior is obtained also for other values of $\gamma,\, \omega$ within or near the gap (see also Section~\ref{sec:Discussion}). This result motivates us to explore in Section~\ref{sec:mod_slfsim} modified self-similar solutions that describe the approach of the flow to self-similarity.

\subsection{Modified self-similar solutions}
\label{sec:mod_slfsim}

To quantitatively consider the approach to self-similarity, we examine solutions of the hydrodynamic equations of the form
\begin{eqnarray}\label{eq:pert definition}
u(r,t)&=&\dot{R}\xi[U(\xi,\delta(t))+\eta U_1(\xi,\delta(t))],\nonumber \\
c(r,t)&=&\dot{R}\xi[C(\xi,\delta(t))+\eta C_1(\xi,\delta(t))], \nonumber \\
\rho(r,t)&=&BR^{\varepsilon}[G(\xi,\delta(t))+\eta G_1(\xi,\delta(t))], \nonumber \\
p(r,t)&=&BR^{\varepsilon}\dot{R}^{2}[P(\xi,\delta(t))+\eta P_1(\xi,\delta(t))].
\end{eqnarray}
Here, $\delta(t)\equiv d\ln\dot{R}/d\ln R$, $\eta\equiv d\delta/d\ln R$, and $F(\xi,\delta(t))$ stand for the self-similar solution $F(\xi)$ obtained for the instantaneous value of $\delta$, $\delta(t)$. The flow fields are of the general form
\begin{equation}\label{eq:general pert}
f(r,t)=R^{\alpha}\dot{R}^{\beta}[F(\xi,\delta(t))+\eta F_1(\xi,\delta(t))].
\end{equation}
For this form,
\begin{eqnarray}\label{eq:general r diff}
\left(\frac{\partial f}{\partial r}\right)_{t}
= R^{\alpha-1}\dot{R}^{\beta}(F'+\eta F_1'),
\end{eqnarray}
where $'\equiv\partial/\partial\xi$, and, using $\dot{\delta}=\eta\dot{R}/R$,
\begin{eqnarray}\label{eq:general t diff}
\left(\frac{\partial f}{\partial t}\right)_{r}=
&\alpha& R^{\alpha-1}\dot{R}^{\beta+1}[F(\xi,\delta(t))+\eta F_1(\xi,\delta(t))] \nonumber \\
&+&\beta R^{\alpha}\dot{R}^{\beta-1}\ddot{R}[F(\xi,\delta(t))+\eta F_1(\xi,\delta(t))] \nonumber \\
&+&R^{\alpha}\dot{R}^{\beta}[(F'+\eta F_1'(\xi,\delta(t)))(-\frac{r}{R^{2}}\dot{R})\nonumber \\
&+&\frac{\partial F}{\partial \delta} \dot{\delta}+\eta\frac{\partial  F_1}{\partial \delta}\dot{\delta}+\dot{\eta} F_1] \nonumber \\
=&&R^{\alpha-1}\dot{R}^{\beta+1}[(F+\eta F_1)(\alpha+\beta\delta)-\xi( F'+\eta F_1')\nonumber\\
&+&\eta(\frac{\partial
F}{\partial\delta}+\eta\frac{\partial  F_1}{\partial\delta}+\frac{d\ln\eta}{d\ln R} F_1)].
\end{eqnarray}

Restricting to solutions with $d\ln\eta/d\ln R=0$ and neglecting $\eta^2$ terms, assuming $\eta\equiv d\delta/d\ln R\ll1$, we obtain a set of ordinary differential equations for $U_1,\,C_1,\,G_1$, which may be written as
\begin{equation}\label{eq:main pert}
\mathbf{A}\left(\begin{array}{c}
  U_1 \\
   G_1 \\
   P_1
\end{array}\right)' = \mathbf{B}\left(\begin{array}{c}
  U_1 \\
   G_1 \\
   P_1
\end{array}\right) + \mathbf{C}\frac{\partial}{\partial\delta}\left(\begin{array}{c}
   U \\
   G \\
   P
\end{array}\right).
\end{equation}
The matrices $\mathbf{A}$, $\mathbf{B}$, and $\mathbf{C}$ are given in Appendix~\ref{sec:mod_eqs}. The equations for $U,\,C,\,G$ are the same self-similar equations as before, given in Appendix~\ref{sec:slfsim_eqns}. The boundary conditions at the shock are set by the Rankine--Hugoniot relations,
which imply
\begin{eqnarray}\label{eq:boundary condition for pert by Hugoniot}
U_1(\delta(t),1)=P_1(\delta(t),1)=G_1(\delta(t),1)=0.
\end{eqnarray}
Note that $\eta$ does not appear in the equations describing the modified solutions, Equation~(\ref{eq:main pert}), and cannot therefore be determined by these equations.

\begin{figure}
\epsscale{1} \plotone{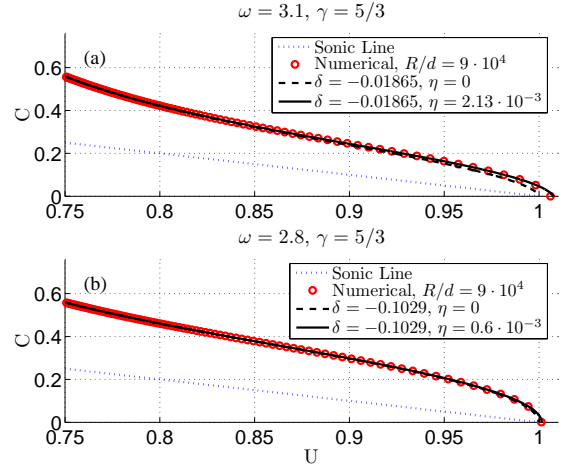} \caption{$C(U)$ curves for explosions with $\omega=3.1, \gamma=5/3$ and $\omega=2.8,\,
\gamma=5/3$. Shown are the numerical solutions, self-similar solutions ($\eta=0$) and modified self-similar solutions ($\eta\neq0$). The difference between the $\eta=0$ and $\eta\neq0$ solutions is difficult to identify in this plot. It is clearly shown in Figures~\ref{CU-pert}--\ref{G-pert}. \label{CU-pert}}
\end{figure}

\begin{figure}
\epsscale{1} \plotone{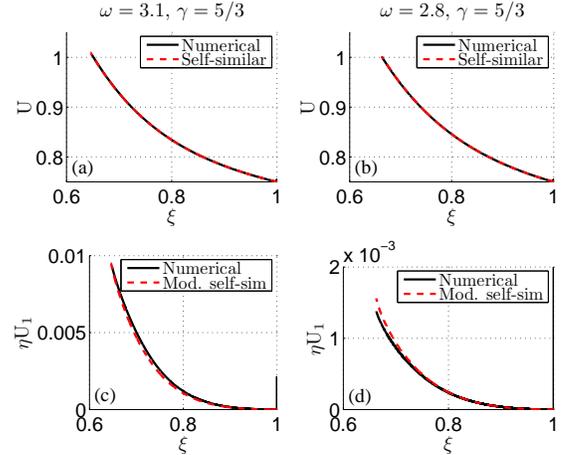} \caption{Numerical and modified self-similar spatial velocity profiles of the calculations presented in Figure~\ref{CU-pert}. Top panels present a comparison of the numerical profiles with the self-similar profiles obtained for the appropriate value of $\delta$ (solutions of Equations~(\ref{eq:dUdC})-(\ref{eq:G})). Bottom panels present a comparison of the difference between the two, i.e., of the numerical deviation from the self-similar solution, with the deviation predicted by the modified self-similar solutions, defined by Equations~(\ref{eq:pert definition}) and determined by Equation~(\ref{eq:main pert}). \label{U-pert}}
\end{figure}

\begin{figure}
\epsscale{1} \plotone{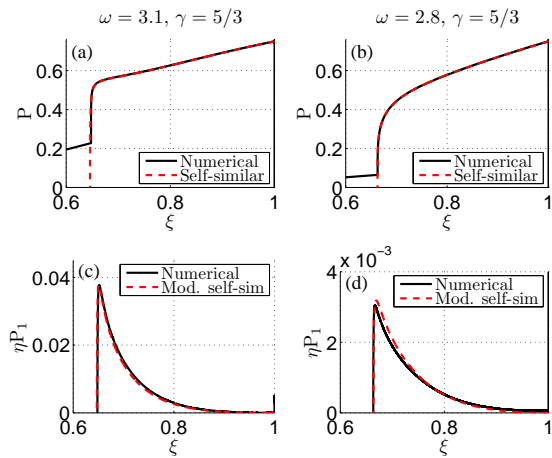} \caption{Same as Figure~\ref{U-pert}, but
for the spatial pressure profiles. \label{P-pert}}
\end{figure}

\begin{figure}
\epsscale{1} \plotone{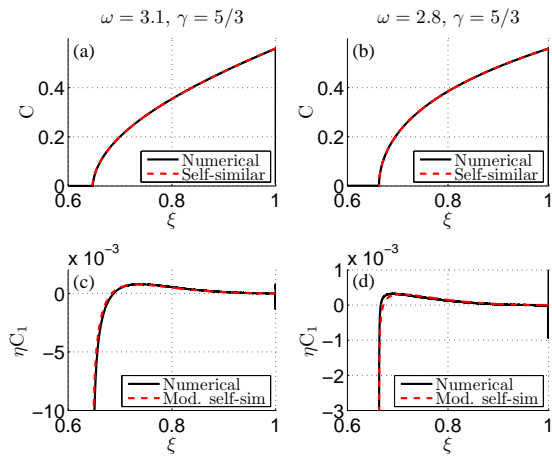} \caption{Same as Figure~\ref{U-pert}, but
for the spatial sound velocity profiles. \label{C-pert}}
\end{figure}

\begin{figure}
\epsscale{1} \plotone{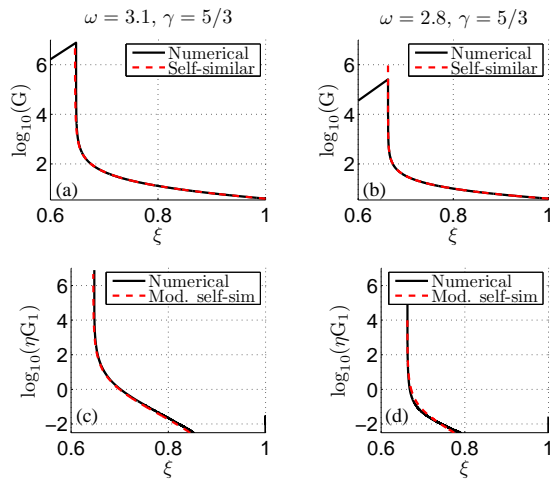} \caption{Same as Figure~\ref{U-pert}, but
for the spatial density profiles. \label{G-pert}}
\end{figure}

We compare in Figures~\ref{CU-pert}--\ref{G-pert} the modified self-similar solutions to the results of numerical calculations, for $\omega=3.1,\, \gamma=5/3$ and $\omega=2.8,\, \gamma=5/3$. We have chosen an example with $\omega=2.8<3$ below the gap, in order to emphasize that slow convergence to self-similarity is not unique to the gap region.
The $C(U)$ curves are compared in Figure~\ref{CU-pert}, and the spatial profiles are compared in Figures~\ref{U-pert}--\ref{G-pert}. The spatial profiles $U_1,\,C_1,\,P_1,\,G_1$ of the numerical solutions were obtained by subtracting from the numerical profiles the self-similar profiles $U,\,C,\,P,\,G$ corresponding to the appropriate value of $\delta(R)$. The value of $\eta$ was determined by comparing the numerical results for $U_1,\,C_1,\,P_1,\,G_1$ with the solutions of Equation~(\ref{eq:main pert}) (note that the solutions of Equation~(\ref{eq:main pert}) are independent of $\eta$, which just determines the normalization of the deviation from the self-similar solution, as given by Equations~(\ref{eq:pert definition})). The agreement between the numerical solutions and the modified self-similar solutions, demonstrated in the plots of Figures \ref{CU-pert}--\ref{G-pert}, suggests that the modified self-similar solutions provide an approximate description of the approach to self-similarity.

\section{Summary and discussion}
\label{sec:Discussion}

We have shown that the self-similar solutions describing the asymptotic flow of the strong explosion problem for $\omega$ values within the gap, $3<\omega<\omega_g(\gamma)$, are the $\delta=0$, $\dot{R}=$~constant, self-similar solutions. For $\omega>3$, the energy in the self-similar solutions is infinite, implying that the self-similar solution may describe the flow only in part of the $(\xi,R)$-plane. This suggests that for $\omega>3$ the asymptotic solution should be composed of a self-similar solution describing the flow at $\xi_c(R)<\xi<1$, matched along some characteristic line $\xi_c(R)$ to a different solution at $0<\xi<\xi_c(R)$, such that the energy contained in the self-similar part of the flow, $\xi_c(R)<\xi<1$, does not diverge as $R\rightarrow\infty$ \citep{WaxmanShvarts93}.

In the analysis of \citep{WaxmanShvarts93}, it was assumed that $\xi_c$ is a $C_+$ characteristic. This was mainly motivated by the fact that requiring the existence of a $C_+$ characteristic, that does not overtake the shock front as $R\rightarrow\infty$, is equivalent to requiring that the solution passes through a sonic point, and it is commonly accepted that the similarity exponents of a second-type solution are determined by the requirement that the solution passes through such a singular point. We showed here that in order to determine the similarity exponents it is sufficient to require the existence of any characteristic, that does not overtake the shock, and for which the energy contained in the self-similar part of the flow, $\xi_c(R)<\xi<1$, does not diverge as $R\rightarrow\infty$.

Examining $\omega$ values below, within, and above the ``gap'', we showed that for each value of $\omega$ there is only one value of $\delta$, $\delta=\delta_*(\omega)$, which yields a valid physical self-similar solution (see Table~\ref{table}). For $\delta<\delta_*$ the $C(U)$ curve determined by Equation~(\ref{eq:dUdC}) crosses the sonic line $U+C=1$ at a non-singular point (yielding a non-single-valued solution, see Equations~(\ref{eq:quadrature}) and~(\ref{eq:deltas})). For $\delta>\delta_*$, the self-similar solution energy diverges, in the sense that the energy contained in $\xi_c(R)<\xi<1$ diverges for any choice of a characteristic, $\xi_c(R)$, that does not overtake the shock wave. The physical solutions are the ST solutions, $\delta_*=\delta_{\textrm{ST}}=(\omega-3)/2$ for $\omega<3$, the solutions derived in \citep{WaxmanShvarts93,WaxmanShvarts10}, $\delta_*=\delta_{\textrm{WS}}<\delta_{\textrm{ST}}$, for $\omega>\omega_g$, and the $\delta_*=0$ solutions for $3<\omega<\omega_g$. The $C(U)$ curves of the $3<\omega<\omega_g$ solutions do not cross the sonic line and $\xi_c(R)$ must be chosen as a $C_0$ characteristic for these solutions.

\begin{figure}
\epsscale{1} \plotone{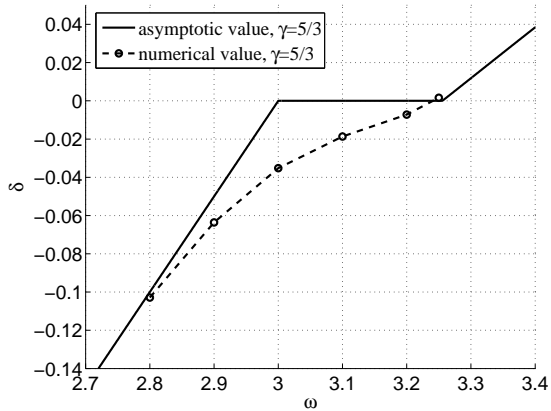} \caption{Self-similar exponent
$\delta$ as a function of $\omega$ for $\gamma=5/3$ (solid line), compared with the value of $\delta$ inferred from numerical simulations (using the method described in Section~\ref{sec:mod_slfsim}) at $R/d=9\times10^4$ (dashed).
\label{delta-omega}}
\end{figure}

In Section~\ref{sec:numerical solution}, we compared the asymptotic, $R/d\gg1$, behavior of numerical solutions of the hydrodynamic equations, Equations~(\ref{eq:hydro_eq}), to that expected based on the $\delta=0$ self-similar solutions. We find that while the flow approaches a self-similar behavior with $|\delta|\ll1$, the convergence to self-similarity is very slow for $\omega\sim3$ (e.g., Figure~\ref{deltaofR}). It should be noted that convergence to self-similarity is slow for any value of $\omega\sim3$, both within and below the gap, as demonstrated in Figure~\ref{delta-omega}. Hence, it is difficult to check using numerical solutions whether for $\omega$ values within the gap the flow indeed approaches a $\delta=0$ self-similar behavior as $R\rightarrow\infty$. We showed in Section~\ref{sec:mod_slfsim} that in this case the flow may be described by a modified self-similar solution, $d\ln\dot{R}/d\ln R=\delta$ with slowly varying $\delta(R)$, $\eta\equiv d\delta/d\ln R\ll1$. In these solutions, the spatial profiles are given by a sum of the self-similar solution corresponding to the instantaneous value of $\delta$ and a self-similar correction linear in $\eta$, see Equations~(\ref{eq:pert definition}). The equations describing the self-similar corrections are given in Equation~(\ref{eq:main pert}). We have shown that for $\omega\sim3$ the modified self-similar solutions provide an approximate description of the flow at large $R$, see Figures~\ref{U-pert}--\ref{G-pert}, with $\delta\rightarrow0$ (and $\eta\neq0$) for $3\leq\omega\leq\omega_{g}$. These results support the conclusion that the flow approaches the $\delta=0$ self-similar solutions as $R$ diverges.

Based on the analysis presented here, we suggest that the definition of first- and second-type similarity solutions should be somewhat modified, and that the family of second-type solutions should be expanded. Solutions of the first-type may be defined as solutions that are valid over the entire $(r,t)$-plane (or the part of which where the flow takes place). Such solutions must satisfy the global conservations laws (of mass, momentum, and energy), and hence the values of the similarity exponents of such solutions may be determined by dimensional considerations. Solutions of the second-type may be defined as solutions, which are valid only in part of the region in the $(r,t)$-plane over which the flow takes place. Such solutions should be required to allow the existence of a characteristic line, $\xi_c(R)$, along which the self-similar solution is matched to another solution, and to comply with the global conservation laws within the region of the $(r,t)$-plane described by the self-similar solution.

\acknowledgments This research was partially supported by ISF, AEC and Minerva grants.



\appendix


\section{The equations describing self-similar flows}
\label{sec:slfsim_eqns}

The equations describing adiabatic one-dimensional flow of an ideal gas are \citep[e.g.][]{LandauLifshitz}
\begin{eqnarray}
\label{eq:hydro_eq}
(\partial_{t}+u\partial_{r})\ln\rho+ r^{-(n-1)}\partial_{r}(r^{n-1}u) &=& 0,
\nonumber \\
(\partial_{t}+u\partial_{r})u+\rho^{-1}\partial_{r}(\gamma^{-1}\rho c^{2}) &=&
0, \nonumber \\
(\partial_{t}+u\partial_{r})(c^{2}\rho^{1-\gamma}) &=& 0,
\end{eqnarray}
where $n=1,2,3$ for planar, cylindrical, and spherical symmetry, respectively.

Substituting Equations~(\ref{eq:ss_scaling}) and~(\ref{eq:Rdot}) in the hydrodynamic Equations~(\ref{eq:hydro_eq}), the partial differential equations are replaced with \citep{ZeldovichRaizer,WaxmanShvarts10} a single ordinary differential equation,
\begin{equation}\label{eq:dUdC}
\frac{dU}{dC}=\frac{\Delta_{1}(U,C)}{\Delta_{2}(U,C)},
\end{equation}
and one quadrature
\begin{equation}\label{eq:quadrature}
\frac{d\ln\xi}{dU}=\frac{\Delta(U,C)}{\Delta_{1}(U,C)}\qquad {\rm
or} \qquad \frac{d\ln\xi}{dC}=\frac{\Delta(U,C)}{\Delta_{2}(U,C)}.
\end{equation}
$G$ is given implicitly by
\begin{equation}\label{eq:G}
(\xi C)^{-2(n+\epsilon)}|1-U|^{\lambda}G^{(\gamma-1)(n+\epsilon)+\lambda}\xi^{n\lambda}={\rm const}
\end{equation}
with
\begin{equation}\label{eq:lambda}
\lambda=-(\gamma-1)\epsilon+2\delta.
\end{equation}
The functions $\Delta$, $\Delta_{1}$, and $\Delta_{2}$ are
\begin{eqnarray}\label{eq:deltas}
\Delta&=&C^{2}-(1-U)^{2}, \nonumber \\
\Delta_{1}&=&U(1-U)(1-U-\delta)-C^{2}\left(n U+\frac{\epsilon+2\delta}{\gamma}\right), \nonumber \\
\Delta_{2}&=&C\{(1-U)(1-U-\delta) \nonumber \\
&-&\frac{\gamma-1}{2}U\left[(n-1)(1-U)+\delta\right]-C^{2} \nonumber \\
&+&\frac{2\delta-(\gamma-1)\epsilon}{2\gamma}\frac{C^{2}}{1-U}\}.
\end{eqnarray}

\section{The equations describing modified self-similar solutions}
\label{sec:mod_eqs}

The matrices $\mathbf{A}$, $\mathbf{B}$, and $\mathbf{C}$ of Equation
~(\ref{eq:main pert}) are given, for spherical symmetry, by
\begin{equation}\label{eq:A diff}
\mathbf{A}=\left(\begin{array}{ccc}
  -\xi G & \xi(1-U) & 0 \\
  \xi^{2}G(1-U) & 0 & -1 \\
  0 & -\gamma\xi P(1-U) & \xi G(1-U)
\end{array}\right),
\end{equation}
\begin{equation}\label{eq:B diff}
\mathbf{B}=\left(\begin{array}{ccc}
  3G+\xi G' & \varepsilon+3U+\xi U' & 0 \\
  \xi G(\delta+2U-1)+\xi^{2} U'G & \xi U(\delta+U-1)+\xi^{2} U'(U-1) & 0 \\
  \xi GP'-\gamma\xi PG' & P(\varepsilon(1-\gamma)+2\delta)+\xi P'(U-1) &
  G(\varepsilon(1-\gamma)+2\delta)-\xi G'\gamma (U-1)
\end{array}\right)
\end{equation}
and
\begin{equation}\label{eq:C diff}
\mathbf{C}=\left(\begin{array}{ccc}
  0 & 1 & 0 \\
  \xi G & 0 & 0 \\
  0 & -\gamma P & G
\end{array}\right).
\end{equation}


\bibliographystyle{hapj}
\bibliography{ms}


\end{document}